\documentclass[sigconf,screen]{acmart}

\setcopyright{rightsretained}
\acmPrice{}
\acmDOI{10.1145/3460319.3464832}
\acmYear{2021}
\copyrightyear{2021}
\acmSubmissionID{issta21main-p77-p}
\acmISBN{978-1-4503-8459-9/21/07}
\acmConference[ISSTA '21]{Proceedings of the 30th ACM SIGSOFT International Symposium on Software Testing and Analysis}{July 11--17, 2021}{Virtual, Denmark}
\acmBooktitle{Proceedings of the 30th ACM SIGSOFT International Symposium on Software Testing and Analysis (ISSTA '21), July 11--17, 2021, Virtual, Denmark}

\usepackage{booktabs}
\usepackage{color}
\usepackage{amsthm}
\usepackage{float}
\usepackage{listings}
\usepackage{algorithm}
\usepackage[noend]{algpseudocode}
\usepackage{graphicx}
\usepackage{courier}
\usepackage{float}
\usepackage{color}
\usepackage{multicol}
\usepackage{multirow}
\usepackage[many]{tcolorbox}
\usepackage{subcaption}
\usepackage{microtype}
\usepackage{syntax}
\usepackage{forest}
\usepackage{framed}

\usepackage{tikz}
\usetikzlibrary{matrix}
\usetikzlibrary{shapes.multipart}
\usetikzlibrary{patterns}
\usetikzlibrary{positioning,fit,calc}
\usetikzlibrary{decorations.pathmorphing}
\usetikzlibrary{decorations.pathreplacing}
\usetikzlibrary{quotes}
\usetikzlibrary{graphs}
\usetikzlibrary{arrows.meta}
\usetikzlibrary{shapes}
\usepackage{smartdiagram}
\usepackage{csvsimple}
\usepackage{multirow}
\usepackage{stackengine,graphicx,scalerel}
\usepackage{soul}
\stackMath

\newcommand\subseq{\sqsubseteq}

\theoremstyle{definition}
\newtheorem{definition}{Definition}
\newtheorem{theorem}{Theorem}

\newcommand{\helium}{{\it Helium }}
\newcommand{\codesonar}{the {\it Commercial Tool} }

\lstdefinestyle{base}{
  language=C,
  emptylines=1,
  breaklines=true,
  aboveskip=0em,
  belowskip=0em,
  basicstyle=\footnotesize\ttfamily\color{black},
  moredelim=**[is][\color{blue}]{@}{@},
  moredelim=**[is][\color{purple}]{~1}{~1},
  moredelim=**[is][\color{brown}]{~2}{~2},
  moredelim=**[is][\color{gray}]{~3}{~3},
  moredelim=**[is][\color{orange}]{~4}{~4},
  moredelim=**[is][\color{violet}]{~5}{~5},
}
\lstdefinestyle{graycode} {
  language=C,
  emptylines=1,
  breaklines=true,
  basicstyle=\footnotesize\ttfamily\color{gray!50},
  moredelim=**[is][\color{blue}]{@}{@},
}

\begin{document}
  \title{Validating Static Warnings via Testing Code Fragments}
  \author{Ashwin Kallingal Joshy}
  \affiliation{%
    \institution{Iowa State University}
    \city{Ames}
    \state{Iowa}
    \country{USA}
  }
  \email{ashwinkj@iastate.edu}
  \author{Xueyuan Chen}
  \affiliation{%
    \institution{Iowa State University}
    \city{Ames}
    \state{Iowa}
    \country{USA}
  }
  \email{xueyuan@iastate.edu}
  \author{Benjamin Steenhoek}
  \affiliation{%
    \institution{Iowa State University}
    \city{Ames}
    \state{Iowa}
    \country{USA}
  }
  \email{benjis@iastate.edu}
  \author{Wei Le}
  \affiliation{%
    \institution{Iowa State University}
    \city{Ames}
    \state{Iowa}
    \country{USA}
  }
  \email{weile@iastate.edu}



\begin{abstract}
  Static analysis is an important approach for finding bugs and vulnerabilities in software. However, inspecting and confirming static warnings are challenging and time-consuming. In this paper, we present a novel solution that automatically generates test cases based on static warnings to validate true and false positives. We designed a syntactic patching algorithm that can generate syntactically valid, semantic preserving executable code fragments from static warnings. We developed a build and testing system to automatically test code fragments using fuzzers, KLEE and Valgrind. We evaluated our techniques using 12 real-world C projects and 1955 warnings from two commercial static analysis tools. We successfully built 68.5\% code fragments and generated 1003 test cases. Through automatic testing, we identified 48 true positives and 27 false positives, and 205 likely false positives. We matched 4 CVE and real-world bugs using {\it Helium}, and they are only triggered by our tool but not other baseline tools. We found that testing code fragments is scalable and useful; it can trigger bugs that testing entire programs or testing procedures failed to trigger.
\end{abstract}

    \maketitle
\section{Introduction}
Static analysis is an important approach to find bugs and vulnerabilities. Path-sensitive static analysis tools~\cite{esp,espx,doulecheck,Klocwork,Ctest,Astree,coverity,SonarQube} are especially useful because it is more precise and reports path information to help diagnose the bugs. Microsoft uses ESP~\cite{esp}, Prefix~\cite{prefix} and ESPx~\cite{espx} to detect resource leaks and buffer overflows. The companies like Grammatech, Synopsys and MathWorks sell their path-sensitive static analysis tools to more than thousands of customers yearly and find bugs in real-world software daily. Although useful, static analysis tools are expensive to use. There can be a large number of warnings generated, and inspecting these messages to confirm true positives and false positives is not a trivial task.

Due to the importance, there have been persistent efforts to help process static warnings.  Flynn et al. developed a classification model that integrated multiple static analysis tools to identify likely true positives~\cite{Flynn:2018}. Zhang et al. developed an interactive approach to learn from the users' feedback to prioritize static warnings~\cite{Zhang:2017:oopsla}. There is also a joint effort from Google and the academia that developed a {\it logistic regression} analysis to predict accurate and actionable static warnings generated from Google software~\cite{ruthruff2008predicting}.

In this paper, we provide a complementary approach by converting the paths reported from static analysis tools to executable test cases. We further leverage advanced testing techniques such as fuzzers and symbolic executors to automatically test the warnings. By demonstrating the symptoms through testing, static warnings can be validated. Different from previous research~\cite{cnc,warningslice:2012} that combines static and dynamic approaches for bugs,  we propose {\it testing code fragments}; that is, we aim to generate test cases that are as small as possible but can still encapsulate the reported buggy paths. We believe that when a test case is small and contains a targeted bug, automatic testing tools can be more scalable and effective to trigger the bug. We can also avoid the potentially complicated setups needed for testing integrated software. Previous research~\cite{warningslice:2012} has mentioned that testing smaller programs can reduce the time of testing and correction of error by developers.

We identified a set of challenges to achieve this goal. First, the code fragments (warning paths, e.g., see Figure~\ref{subfig:statWar}) reported by static warnings typically are not compilable. We need to patch the code fragments to fix the syntax errors. When adding the patch to generate a test case, we should not alter the paths reported in the warnings. Second, we need to isolate the dependencies from the project required to build the code fragments. Third, to enable automation, we need to provide proper test harnesses and inputs that can trigger the symptoms aligned with the warning types. Additionally, we need to construct test oracles that can match the symptoms reported from testing and static warnings.

We developed the {\it Helium} framework to address these challenges. It consists of a {\it {Lowest Common Ancestor (LCA)}-based syntactic patching} algorithm and a system for building and testing code fragments. We formally define what {\it syntactic patches} we should generate and prove that following this definition, the warning paths in the original program will be preserved in the test case we generated. Different from the existing techniques of auto-fixing compiler errors~\cite{1963-CACM-Error-Irons,1972-Journal-Aho-Minimum,1973-POPL-Graham-Practical,2019-DeepReinforce,2017-DeepFix}, our approach analyzes the parse trees to preserve the semantics of code fragments and also make sure the patch is as small as possible.

To resolve dependencies, we analyzed the code fragment and its surrounding environment in the original project to determine the header files, the definitions of functions, types and variables, compiler flags, and the libraries needed to compile and link the code fragments.

For testing, we used randomly generated test inputs as well as the automatically generated test inputs through fuzzers like {\it Radamsa}~\footnote{https://gitlab.com/akihe/radamsa} and the symbolic executors like {\it KLEE}~\footnote{https://klee.github.io/}. Based on testing results, we identify {\it valid} tests to confirm true positives and false positives. The valid tests match the failures specified in our test oracles. We constructed test oracles using static warnings,  including failure locations, e.g., the file names and line numbers, as well as failure symptoms linked to the warning types, e.g., buffer overflows, null-pointer dereferences, and memory leaks. We used {\it Valgrind}~\footnote{https://valgrind.org/} and also added {\it assert} statements to capture such failures during testing.

We implemented {\it Helium} and evaluated our techniques using 12 C projects such as {\tt cvs}, {\tt httpd}, {\tt findutils}, {\tt grep}, {\tt make} and {\tt coreutils}. We collected 1955 static warnings from  two popular commercial static analysis tools, {\it PolySpace} from MathWorks and the {\it Commercial Tool} (anonymized based on our license). We used  Radamsa, KLEE and Valgrind as our automatic testing tools. \helium compiled 1340 code fragments and generated {1003} test cases. We identified {48} true positives and {27} false positives, and categorized {205} as likely false positives. Our techniques significantly outperformed the baselines, including unit testing, {\it RLAssist}~\cite{2019-DeepReinforce}, {\it MACER}~~\cite{2020-Macer}, {\it BovInpsector}~\cite{2015-ASE-BovInspector} and the integrated testing using existing test suites. We matched 4 true positives to CVE and real-world bugs that the other tools did not find.

In this paper, we make the following contributions: 
\begin{enumerate}
  \item defining {\it syntactic patching} for code fragments (\S 3),
  \item designing an algorithm that can generate syntactically valid, semantic-preserving, as small as possible code fragments (\S 4),
  \item developing a system that automatically builds and tests code fragments (\S 5); and
  \item building a tool, {\it Helium}, using which, we demonstrated that our techniques can effectively validate static warnings from real-world code and static analysis tools (\S 6).
\end{enumerate}

Our paper website is {\url{https://sites.google.com/view/helium-2021}}.

\section{Overview}
We motivate our work using a real-world example. We then explain at a high level how \helium works to help developers. 

\subsection{A Motivating Example}~\label{subsec:Motivate}
In Figure~\ref{subfig:statWar}, we present a static warning reported by \codesonar for {\tt squid-2.3}. The warning includes a list of statements that lead to a buffer overflow at line~1025 in {\tt ftp.c}. These statements are located in three different functions and files. To confirm the warning, the developer need to inspect the path and understand what happens along the path. The task is difficult and time consuming, considering there are hundreds of warnings reported for {\tt squid-2.3} by the {\it Commercial Tool}.

   \definecolor{codeNumbers}    {rgb}{0.0, 0.0, 0.0}
  \definecolor{celadon}{rgb}{0.67, 0.88, 0.69}
  \definecolor{cardinal}{rgb}{0.77, 0.12, 0.23}
  \definecolor{darkpastelred}{rgb}{0.76, 0.23, 0.13}
  \lstset{%
    frame            = b,                               
    tabsize          = 1,                               
    numbers          = left,                            
    framesep         = 0.4pt,                           
    framerule        = 0.4pt,                           
    showstringspaces = false,                           
    language         = C++,                             
    escapechar       = \!,                              
    breaklines       = true,                            
    captionpos       = b,                               
    keepspaces       = true,                            
    numbers          = left,                            
    numbersep        = 4pt,                             
    numberstyle      = \footnotesize\color{codeNumbers},  
    stepnumber       = 1,                               
  }
  \lstset{escapechar=^}
  \begin{figure*}
    \centering
    \begin{minipage}[b]{0.48\textwidth}
      \begin{subfigure}[b]{\textwidth}
        \centering
\begin{lstlisting}[frame=none,numbers=none]
file:lib/rfc1738.c
89: char *rfc1738_do_escape(char *url, int encode_reserved) {
102: for (p = url, q = buf; *p != 0; p++) {
103:   do_escape = 1;
107:   if (*p == rfc1738_reserved_chars[i]) {
108:     do_escape = 1;
109:     break;
139:   if (do_escape == 1) {
140:     (void) sprintf(q, "%%%02x", (unsigned char) *p);
141:     q += sizeof(char) * 2;
142:   } else {
143:     *q = *p;
146:  *q = '\0';
147:  return (buf);



file:lib/rfc1738.c
175: char *rfc1738_escape_part(char *url) {
177:  return rfc1738_do_escape(url, 1);



file:src/ftp.c
999:void ftpBuildTitleUrl(FtpStateData *ftpState) {
1004:  len = 64 + strlen(ftpState->user)
1005:    + strlen(ftpState->password)
1006:    + strlen(request->host)
1007:    + strLen(request->urlpath);
1008:  t = ftpState->base_href = xcalloc(len, 1);
1020:  strcat(t, "ftp://");
1021:  if (strcmp(ftpState->user, "anonymous")) {
1022:    strcat(t, rfc1738_escape_part(ftpState->user));
1023:    if (strcmp(ftpState->user, "anonymous")) {
1024:      strcat(t, ftpState->user);
1025:      strcat(t, "@");




\end{lstlisting}
        \caption{static warning for {\tt squid-2.3}}\label{subfig:statWar}
      \end{subfigure}

            \begin{subfigure}[b]{\textwidth}
        \centering
\begin{lstlisting}[frame=none,numbers=none]

./squid \ \ \ \ \ \ \ \ \ \ \ \ \ \ \ \ \ \ \ \ \ \ \ \ \ \ \ \ \ \ \ \ \ \ \ \ \ \ \ \ \ \ \ \ \ \ \ \ \ \ \ \ \ \ \ \ \ \ \ \ \ \ \ \ \ \ \ \ \  password@host.com

\end{lstlisting}
        \caption{testing code fragments using the input provided by {\tt squid-2.3}}\label{subfig:input}
      \end{subfigure} 

    \end{minipage}%
    \qquad
    \begin{minipage}[b]{0.48\textwidth}
      \begin{subfigure}[b]{\textwidth}
        \centering
        \sethlcolor{yellow}
\begin{lstlisting}[frame=none]
^\hl{\#include <stdlib.h>}^
^\hl{\#include <squid.h>}^ 
...
^\hl{void *xcalloc(int n, size_t sz);}^
^\hl{typedef struct _Ftpdata \{}^
^\hl{    char user[MAX_URL];}^
^\hl{    char password[MAX_URL];   ...}^
^\hl{\} FtpStateData;}^
\end{lstlisting}
        \vspace*{-6pt}
        \sethlcolor{red}
\begin{lstlisting}[frame=none,firstnumber=9]
char *rfc1738_do_escape(char *url, int encode_reserved) {
  for (p = url, q = buf; *p != 0; p++) {
    ^\hl{for (i = 0; i < sizeof((rfc1738_reserved_chars)) \&\& encode_reserved > 0; i++) \{}^
      if (*p == rfc1738_reserved_chars[i]) {
        do_escape = 1;
        break; ^\hl{\}}^
      if (do_escape == 1) {
        (void) sprintf(q, "%%%02x", (unsigned char) *p);
        q += sizeof(char) * 2;
      } else {
        *q = *p; ^\hl{\}}^  ^\hl{\}}^
  *q = '\0';
  return (buf); ^\hl{\}}^
char *rfc1738_escape_part(char *url) {
  return rfc1738_do_escape(url, 1); ^\hl{\}}^
void ftpBuildTitleUrl(FtpStateData *ftpState) {
  request_t *request = ftpState->request;
\end{lstlisting}
        \sethlcolor{yellow}
        \vspace*{-6pt}
\begin{lstlisting}[frame=none,firstnumber=26]
  ^\hl{size_t len; char *t;}^
\end{lstlisting}
        \vspace*{-6pt}
        \sethlcolor{red}
\begin{lstlisting}[frame=none,firstnumber=27]
  len = 64 + strlen(ftpState->user)
    + strlen(ftpState->password) + strlen(request->host)
    + strLen(request->urlpath);
  t = ftpState->base_href = xcalloc(len, 1);
  strcat(t, "ftp://");
  if (strcmp(ftpState->user, "anonymous")) {
    strcat(t, rfc1738_escape_part(ftpState->user));
    if (strcmp(ftpState->user, "anonymous")) {
      strcat(t, ftpState->user);
      strcat(t, "@"); ^\hl{\}}^ ^\hl{\}}^ ^\hl{\}}^
\end{lstlisting}
        \vspace*{-6pt}
        \sethlcolor{celadon}
\begin{lstlisting}[frame=none,firstnumber=37]
^\hl{int main(int argc, char *argv[])\{}^
  ^\hl{FtpStateData *ftpState=xcalloc(1,sizeof(FtpStateData));}^
  ^\hl{ftpState->user = argv[1];}^
  ^\hl{ftpState->password = argv[2];  $\dots$}^
  ^\hl{ftpBuildTitleUrl(ftpState); \}}^
\end{lstlisting}
        \caption{test case generated by \helium }\label{subfig:heliumOut}
      \end{subfigure}
    \end{minipage}
    \caption{the buffer overflow example in {\tt squid-2.3} relevant to CVE 2002--0068~\label{example}}
  \end{figure*}

Using \helium, we generate an executable test case shown in Figure~\ref{subfig:heliumOut}. Specifically, \helium adds the red lines (lines~11, 14, 19, 21, 23 and 36) to make the code syntactically valid.  Note that the code may be parsable by adding just "{\tt\}}"s; however, without adding line~11 (this loophead exists in the original program but \codesonar did not include it in the static warning), the test case cannot preserve the semantics of the original program, as {\tt break} at line~14 would be paired with the loop at line~10 instead of the one at line~11.

In addition, \helium adds the yellow lines to resolve the dependencies needed to build the code fragment, including variables (line~26), type (lines~5--8) and function (line~4) dependencies as well as the header files (lines~1--3). \helium also generates the test harness code at lines~37-41 (shown in green).  Running the test input shown in Figure~\ref{subfig:input}, \helium triggered the buffer overflow (similar to CVE 2002--0068). We thus can confirm that this warning is a true positive.

\subsection{How Helium Works to Help the Developers}~\label{subsec:HeliumFrame}
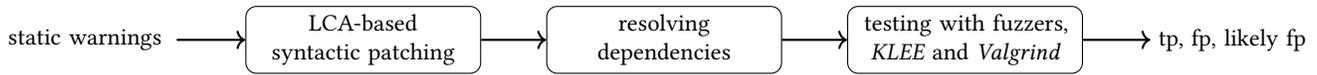
\begin{figure*}
  \tikzstyle{block} = [rectangle,draw,text width=9.2em, text centered, rounded 
  corners, minimum height=2em]
  \tikzstyle{io} = [rectangle, text width = 8.0em, align=center]
  \tikzstyle{io2} = [rectangle, text width = 7.0em, align=center]
  \tikzstyle{output} = [rectangle, text width = 9.0em, align=left]
  \tikzstyle{bigbox} = [rectangle,draw,text width=11em, text centered, rounded 
  corners, minimum height=10em]
  \tikzstyle{bigbox2} = [rectangle,draw,text width=11em, text centered, rounded 
  corners, minimum height=7.4em]
  \tikzstyle{line} = [draw, -latex']
  \begin{tikzpicture}[node distance = 2.5cm, auto]
    \node[io2](pseudocode){static warnings};
    \node[block, right of=pseudocode,node distance= 3.7cm](pcode){LCA-based \\syntactic patching};
    \node[block, right of=pcode,node distance= 4cm](test){resolving\\ dependencies};
    \node[block, right of=test,node distance= 4cm](random){testing with fuzzers, {\it KLEE} and {\it Valgrind}};
    \node[output, right of=random, node distance=4cm](rank){tp, fp, likely fp};
    \draw [thick,->] (pseudocode) -- (pcode);
    \draw [thick,->] (pcode) -- node[auto]{} (test);
    \draw [thick,->] (test) -- node[auto]{} (random);
    \draw [thick,->](random) -- (rank);
  \end{tikzpicture}
  \caption{The Helium Workflow}
  \label{fig:flowChart}
\end{figure*}

Figure~\ref{fig:flowChart} (in page 4) presents an overview of our approach. \helium takes static warnings as input. It first applies our {\it LCA-based syntactic patching} algorithm to generate a parsable, semantic-preserving code fragment from the warning. It then resolves the dependencies and builds the code fragment to generate an executable test case. Finally, \helium adds assertions and tests code fragments with fuzzers, KLEE and Valgrind. Comparing the test results with our test oracles, it validates whether a warning is a true positive or a false positive. If many tests exercise the paths but do not trigger any failures, we report the warnings as likely false positives. Based on this information, developers then can prioritize which warnings to fix first, and use our test cases to help diagnose the warnings.

When manually inspecting a static warning, the developer aims to understand whether there indeed exists a bug along the paths reported and to predict what consequences this bug may cause. \helium automates this manual process by running tests through the paths.  When reporting a buggy path, static analysis tools had concluded that the path sufficiently implied the error condition, and some (or all) user inputs can trigger the bug. Our testing finds such user inputs, confirms that the error condition is indeed correctly computed by static analysis, and demonstrates its failure symptoms.

\section{Defining Syntactic Patching}~\label{sec:def}
By {\it syntactic patching}, we mean patching a code fragment and making it syntactically valid. The goal of our syntactic patching is to generate a small, syntactically valid, and semantic preserving code fragment. In the following, we provide the details on what this goal means.

\subsection{Code Fragments}
In this paper, we use {\it code fragments} and {\it code segments} interchangeably. We use {\it tokens} to define code fragments. Precisely, the {\it token} here means its {\it lexeme}~\cite{1986-Aho-Dragon}, e.g., the variable ``pi" is a lexeme, and the constant ``3.1415" is also a lexeme. Each token has a unique identification given by its location in the program.

\begin{definition}
  A {\it program}, $p$, is a finite sequence of tokens, $p=t_0 t_1 \ldots t_n$. We use the notation $|p|$ to indicate the length of the program in terms of the number of tokens. We say $p_1$ is {\it smaller} than $p_2$ iff $|p_1|<|p_2|$.

\end{definition}

\begin{definition}
  Given programs $p=t_0 t_1 \ldots t_n$ and $q=t'_0 t'_1 \ldots t'_m$, we say $q$ is a {\it code fragment} of $p$ iff $t'_0 t'_1 \ldots t'_m$ is a subsequence~\cite{1986-Aho-Dragon} of $t_0 t_1 \ldots t_n$, denoted by $q \subseq p$.
\end{definition}

\subsection{LCA-Based Syntactic Patching}~\label{deflca}
We present three different syntactic patching approaches and show that this problem is challenging: {\it token-based syntactic patching} may change semantics and  {\it tree-based} can generate a large patched program. Our novel solution is called {\it LCA-based syntactic patching}.

\begin{definition} In {\it token-based syntactic patching}, given a context-free grammar $G$,  a program $p$, and a code fragment $s$, $s\subseq p$, we say $s'$ is a patched program of $s$ regarding $p$, iff
  \begin{enumerate}
  \item $s' \subseq p$;
  \item $s \subseq s'$;
  \item $s'$ is recognized by $G$; and 
  \item $s'$ is the smallest program that satisfies the conditions (1)--(3).
  \end{enumerate}
\label{def:token}
\end{definition}

This definition is problematic, as demonstrated in Figure~\ref{def:token}: when we select {\tt foo} and {\tt b} on the left, or select {\tt while}, {\tt c2} and {\tt s1} on the right, we generate a patched program (see the bottom) that has a different syntax tree from the original program, which can lead to different semantics.

\begin{figure}[H]
  \centering
  \includegraphics[scale=0.7]{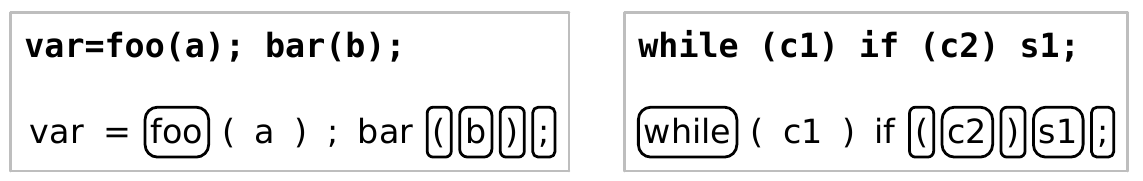}
  \caption{\label{fig:ill-code} the top line is the original code, and the boxed nodes at the bottom represent the patched code under Definition~\ref{def:token}.}
\end{figure}

\begin{definition} In {\it tree-based syntactic patching}, given a context-free grammar $G$,  a program $p$, and a code fragment $s$, $s\subseq p$, we say $s'$ is a patched program of $s$ regarding $p$, iff
  \begin{enumerate}
    \item $s' \subseq p$,
    \item $s \subseq s'$,
    \item $s'$ is recognized by $G$,
    \item $\exists t$, where $t$ is a common subtree of $parse(s', G)$ and
    $parse(p, G)$ such that $s \subseq dft(t)$, and
    \item  $s'$ is the smallest program that satisfies the conditions (1)--(4).
  \end{enumerate}
  \label{def:tree-based}
\end{definition}

Here, \textit{parse(p,G)} returns a parse tree from a given program $p$ and a grammar $G$.
{\textit{dft(t)}}~\label{def:dft} (depth first traversal) produces a program by sequencing the leaf tokens during traversing the parse tree, $t$, in a depth-first order~\cite{1986-Aho-Dragon}.

\begin{figure}[H]
  \centering
  \includegraphics[scale=0.65]{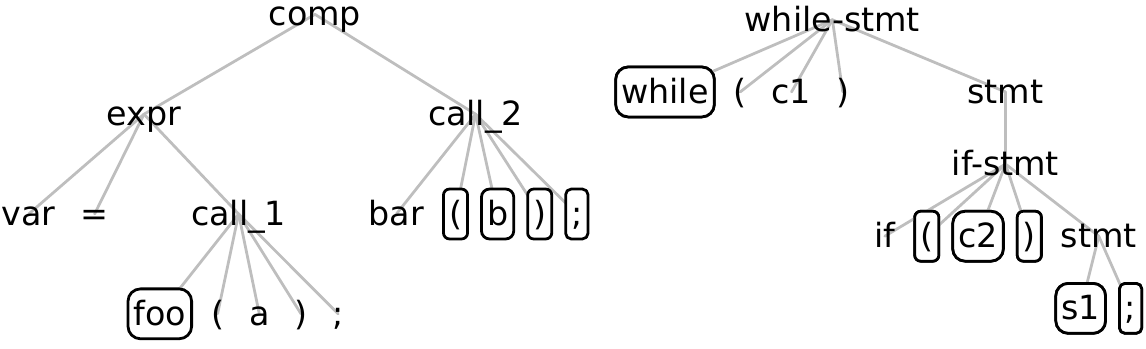}
  \caption{\label{fig:ill-code-resolve} Addressing the problems in Figure~\ref{fig:ill-code}}
\end{figure}

Definition 4 addresses the problem of Definition 3, as shown in Figure~\ref{fig:ill-code-resolve}. On the left, the smallest subtree that contains the selected tokens {\tt foo} and {\tt b} is the entire tree rooted by {\tt comp}. The patched program thus is {\tt var = foo(a); bar(b);} instead of {\tt foo(b)}. Similarly, on the right, when we keep the smallest subtree of the selected nodes {\tt while}, {\tt c2} and {\tt s1}, we include {\tt while(c1)} in the patched program, and {\tt c2}  will not be mistakenly used as a condition for the {\tt while} loop.

\begin{figure}[H]
  \centering \includegraphics[scale=0.8]{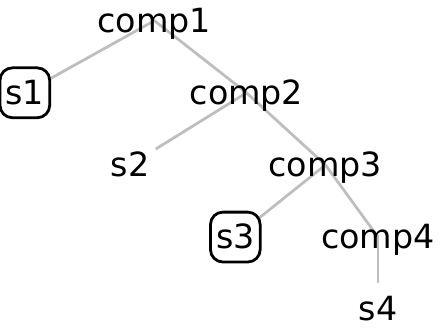}
  \caption{\label{fig:tree-problem} Generating a large patched program}
\end{figure}

Definition 4 is still not ideal as it can lead to an unnecessarily large patched program. For example, in Figure~\ref{fig:tree-problem}, given the selected {\tt s1} and {\tt s3}, we would start from the root {\tt comp1} and generate the patched program {\tt s1}; {\tt s2}; {\tt s3}; {\tt s4}. But without {\tt s2} and {\tt s4}, the program {\tt s1}; {\tt s3} is still compilable and retains the execution order for {\tt s1} and {\tt s3} as in the original program.

 To address the problems of Definitions 3 and 4, we propose {\it LCA-based syntactic patching}. This approach aims to keep the syntactic structure, thus semantics, of code fragments during patching while maintaining a smallest parse tree. The idea is to only copy the important non-terminal nodes that can maintain the syntactic relations of any two selected tokens in the parse tree. We identified that such ``bridge nodes'' are the \textit{Lowest Common Ancestors (LCAs)}~\cite{schieber1988finding}. The LCA of $x$ and $y$ is the common ancestor of  $x$ and $y$ that is not the ancestor of any other common ancestors of $x$ and $y$. 

\begin{definition}
An {\it LCA relation} of two nodes $x$ and $y$ on a parse tree $pt$, denoted by $R_{lca}(x, y, pt)$, is a 4-tuple $(L, r, i_x, i_y)$,
 where
  \begin{enumerate}
  \item $L$ is the LCA of $x$ and $y$, a non-terminal node on the parse tree $pt$,
  \item $r$ is the right hand side (RHS) of the production rule, where $L$ is the left hand side (LHS),
  \item $i_x$ is the position of the $x$'s ancestor in $r$, and
  \item $i_y$ is the position of the $y$'s ancestor in $r$.
  \end{enumerate}
\label{def:lca}
\end{definition}

\begin{figure}[H]
  \centering
  \includegraphics[scale=0.7]{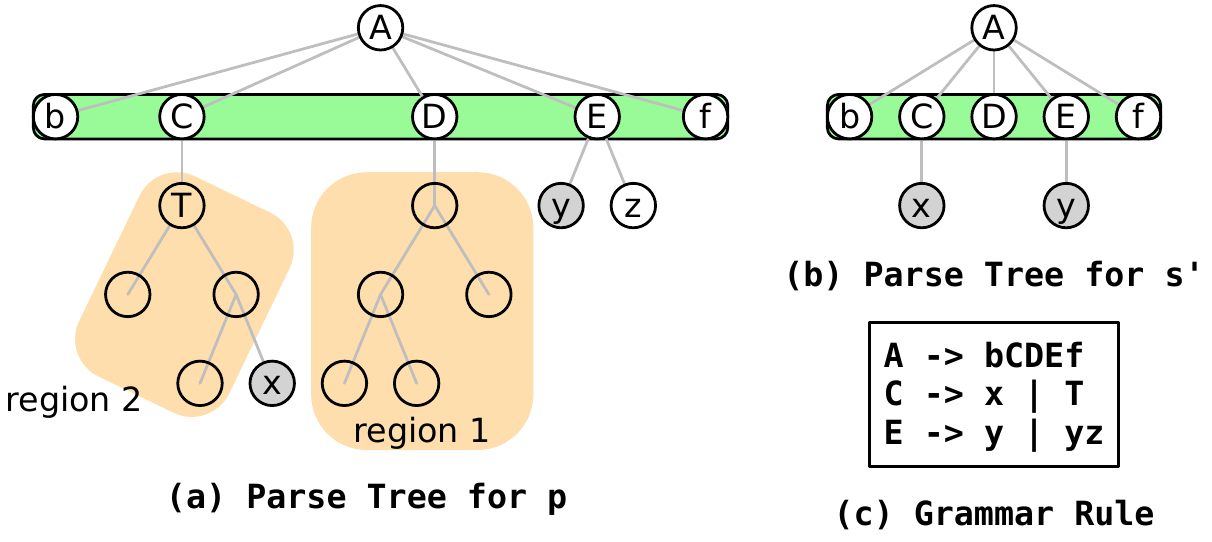}
  \caption{\label{fig:lca-example} An example of the LCA relation}
\end{figure}

In Figure~\ref{fig:lca-example} (a), $A$ is the LCA of the two leaf nodes $x$ and $y$; $r$ is $bCDEf$, the RHS of $A$; the ancestors of $x$ and $y$ in the sequence $bCDEf$ are $C$ and $E$ respectively, and thus $i_x=2$ and $i_y=4$. Therefore, in this example, $R_{lca}(x,y, pt)=(A,bCDEf,2,4)$. This relation captures the production rule used from the LCA node $A$ as well as the nonterminals $C$ and $E$ that will eventually derive to the selected tokens $x$ and $y$.

\begin{definition} In {\it LCA-based syntactic patching}, given a context-free grammar $G$,  a program $p$, and a code fragment $s$, $s\subseq p$, we say $s'$ is a patched program of $s$ regarding $p$, iff
  \begin{enumerate}
    \item $s' \subseq p$,
    \item $s \subseq s'$,
    \item $s'$ is recognized by $G$,
    \item $\forall x,y \in s$, \\
    $R_{lca}(x,y, parse(p, G)) = R_{lca}(x,y, parse(s', G))$, and 
    \item $s'$ is the smallest program that satisfies the conditions (1)--(4).
  \end{enumerate}
  \label{def:pt}
\end{definition}

Condition (4) says that the LCA relation of any two tokens should remain unchanged when generating $s'$ from $p$. Unlike Definition 4, which keeps the entire subtree rooted from the common ancestor of tokens, here we only copy the "first level" of the subtree, including the LCA and its production rule, for generating a small $s'$. In Figure~\ref{fig:lca-example}~(b), the parse tree of the patched program  $s'$ keeps $A$, the LCA of $x$ and $y$, and its production rule $A\rightarrow bCDEf$. When generating patches, instead of using the original product rules for $C$ and $E$, we use $C\rightarrow x$ and $E\rightarrow y$ to keep the patch small.

\subsection{Preserving Semantics}~\label{sec:semantic}
When performing syntactic patching, we need to {\it preserve semantics}, which means that the paths in static warnings should be retained even with the patch, and hence the bug in the warning can be reproduced. We developed two theorems and used the concept of {\it partial order}~\cite{1970-Sigplan-Allen-Control} to demonstrate that our LCA syntactic patching retains the partial order of any two selected tokens; as a result, the execution order of statements along the warning paths are the same in the patched program and in the original program. The proof of Theorem 1 considers {\it sequential}, {\it loop} and {\it branch}, three types of control flow and their nested cases. The proof of Theorem~2 used the Theorem~1 as well as the definitions of partial order and subsequence. Due to the space, we moved the proof sketches here~\footnote{ {\url{https://sites.google.com/view/helium-2021}}}.

\begin{definition}
{\it partial order} is defined on two nodes on the control flow graph. We say node $x_1$ is ordered before $x_2$ ($x_1<x_2$) iff along some acyclic path, $x_1$ is a predecessor of $x_2$.  If $x_1$ and $x_2$ do not appear together on any acyclic path, they do not have an order.
\label{def:partial}
\end{definition}

\begin{theorem}
  If $s'$ is a patched program of the code fragment $s$ regarding the original program $p$ via LCA-based syntactic patching, two statements $n_1$ and $n_2$ in $s'$ have the same partial order in $s$ and $p$. 
\end{theorem}

\begin{theorem}
  If two statements $n_1$ and $n_2$ in $s'$ have the same partial order in $p$, any path in $s'$ is a subsequence of some path in $p$, i.e., $\forall $ path $t$ in $s'$, $\exists$ path $\omega$ in $p$, such that $t\subseq \omega$.
  ~\label{theorem}
\end{theorem}

\section{Computing Syntactic Patches}
We developed an LCA-based syntactic patching algorithm, consisting of two steps: \textit{preserving LCA relations} (Section~\ref{sec:alg}) and \textit{generating minimal patches} (Section~\ref{sec:patch}).

\subsection{Preserving LCA Relations}~\label{sec:alg}
In Algorithm \ref{alg:grammar-patch} at line~4, $N$ stores the parse tree nodes whose LCAs need to be computed. Initially, it is set to $s$.  $\Delta s$ stores the patch in progress. Lines 5--10 present the key step of identifying LCA nodes for the selected tokens. At lines~5--~6, we use a worklist to traverse the nonterminal nodes from the bottom of the parse tree $pt$. For any such node $l$, we check if its descendants overlap with the nodes in $N$. If we found more than two of such nodes, denoted by the set $C_{lca}$ at line~8, we add $l$ to $N$ ($l$ is an LCA) and remove $C_{lca}$ from $N$ at line~10. As the algorithm progresses through the loop, the tokens in $N$ are gradually replaced by their LCAs until the most top level LCA is reached, and the worklist at line~6 becomes empty.

At lines~11-15, we generate the patch based on the LCA relation discovered above. At line~12, when the children of the LCA node are terminals, we directly add them to the patch based on the parse tree, denoted by $\Delta s \gets \Delta s \cup c$. If $c$ is a nonterminal, we use {\sc GenMinPatch} at line~15 to find a minimum patch that derives from $c$ to $target$.  Using the same {\sc GenMinPatch}, at line~16 when $worklist$ is empty, we find a desired derivation from the start symbol, the parse tree root, to the top LCA (the last element processed by $worklist$) to generate the patched program recognizable by the grammar.

\begin{algorithm}
  \caption{LCA-based Syntactic Patching\label{alg:grammar-patch}}
  \begin{algorithmic}[1]
    \State INPUT: $p$ (program), $s$ (code fragment), $G$ (grammar)
    \State OUTPUT: $s'$ (patched program)
    \Statex
    \State $pt \gets$ \Call{Parse}{$p$} \label{line:pt}
    \State $N \gets s, \Delta s \gets \emptyset$
    \State $worklist \gets$ \Call{SortByLevel}{$pt.nonterminals$}
    \While {$worklist \neq \emptyset$}
    \State remove $l$ from $worklist$ \label{line:lowest}
    \State $C_{lca} = l.descendants \cap N$
    \If {$|C_{lca}| \ge 2$} 
    \State $ N \gets  (N \setminus C_{lca}) \cup l$
    \ForAll {$c \in l.children$}
    \If {$c$ is terminal} $\Delta s \gets \Delta s \cup c$
    \Else
    \State $target \gets c.descendants \cap C_{lca}$
    \State $s' \gets s'\cup $
    \Call{GenMinPatch}{$c$, $target$}
    \EndIf
    \EndFor
    \EndIf
    \EndWhile
    \State $s'=s'\cup$ \Call{GenMinPatch}{$pt.root$, $N$}
  \end{algorithmic}
\end{algorithm}

\noindent{\bf Example:} In Figure~\ref{fig:lca-example} (a), when traversing the parse tree in a bottom up fashion, we determine that $A$ is the LCA for the selected tokens $x$ and $y$ (at lines 8--9 in Algorithm~\ref{alg:grammar-patch}, $A$ is $l$, and $x$ and $y$ are in $N$). We add $b$ and $f$ to the patch since they are terminals (line~12). For the nonterminals $C$, $D$ and $E$, we use {\sc GenMinPatch} to find $C\rightarrow x$ and $E\rightarrow y$ that can generate a minimal patch (lines 14 and 15). If $A$ is not the root, we also use {\sc GenMinPatch} to generate a patch that derives the root to $A$ (line 16).

\subsection{Generating Minimal Patches}~\label{sec:patch}
{\sc GenMinPatch} in Algorithm~\ref{alg:grammar-patch} aims to find a {\it shortest derivation} from a nonterminal to a target, defined as follows:

\begin{definition}
  Given a nonterminal $X$, a target $y$ (it can be either terminal or nonterminal), and a string $\alpha$ which derives from $X$ and consists of $y$ and terminals, a {\it shortest derivation} from $X$ to $y$ regarding $\alpha$ generates the string $\beta$, such that (1) $\beta\subseq \alpha$, (2) $y \in \beta$, (3) $\beta$ can also be derived from $X$, and (4) $\beta$ is the shortest string that satisfies the above conditions.
  \label{def}
\end{definition}

To compute the shortest derivation, we define \textit{derivation graph} and use it to reduce this problem to finding the shortest paths on the derivation graph.

\begin{definition}
 A \textit{derivation graph} regarding a context-free grammar and its nonterminal $X$ is a directed graph $G=(V,E)$, where the node represents either a \textit{nonterminal} or the RHS of a production rule. Correspondingly, an edge is either (1) a {\it Type I} edge from a nonterminal to its RHS, and the edge is weighted with the number of terminals used in RHS; or (2) a {\it Type II} edge from RHS to a nonterminal used in the RHS, and the edge does not have a weight. 

\end{definition}

\setlength{\grammarparsep}{1pt plus 1pt minus 1pt}
\setlength{\grammarindent}{4em}

\begin{figure}[H]
  \centering
  \begin{tikzpicture}[every edge quotes/.style={red,auto}]
    \matrix[anchor=north] {
      \node (bnf2) [text width=3.6cm,
        ] {
          \begin{lstlisting}[basicstyle=\footnotesize\ttfamily\bfseries]
      X -> mY|M|uCDv
      Y -> dZf
      Z -> M|k
      M -> mn
      C -> m
      D -> gnh|nd
          \end{lstlisting}
        };&

        \graph [grow right=2.6em] {
          X -> {
            mY [> "1"] ->[dotted] Y -> ["2"] dZf ->[dotted] Z -> ["1"] k,
            M [> "0" right] -> ["2"'] mn,
            uCDv [> "2"'] ->[dotted] {
              C -> ["1"] m,
              D -> ["3"] gnh,
              D -> ["2"] nd
            }
          }
        };
      \draw [->] (Z) -- node [auto,red,right] {0} (M); \\
        \node [font=\footnotesize] {(a) Grammar};&
          \node [font=\footnotesize, shift={(2cm,0)}] {(b) Derivation Graph for $X$};
          \\
        };

    \end{tikzpicture}
    \caption{An example of derivation graph \label{fig:lazy-eval-example}}
  \end{figure}
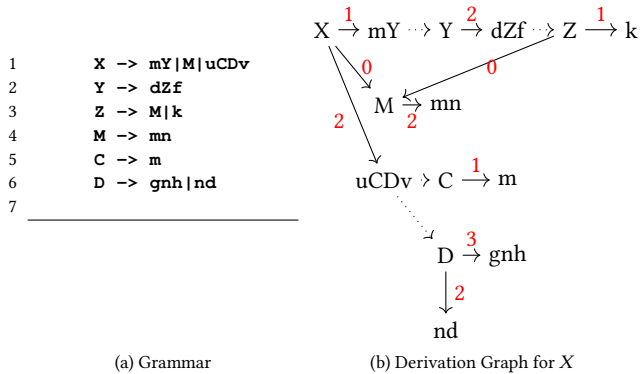

\noindent{\bf Example}: In Figure \ref{fig:lazy-eval-example}, the solid edges from $X$ are the Type I edges. The weights on the edges starting from $X$ indicate that the RHSs of $X$ used 1, 0 and 2 terminals respectively. Some example Type II edges (dashed) are from $uCDv$ to $C$, and from $uCDv$ to $D$.

\begin{algorithm}[ht]
  \caption{Generating Minimal Patches\label{alg:shortest}}
  \begin{algorithmic}[1]
    \State Input: $X$ (nonterminal), $g$ (derivation graph of $X$), $y$ (target) 
    \State Output: \Call{D$_{min}$}{$X$,$y$}
    \Statex
    \Procedure{D$_{min}$}{$X$,$y$}
    \State \Return Min(\Call{D$_{min}$-I}{$X$,$y$}, \Call{D$_{min}$-II}{$X$,$y$})
    \EndProcedure
    \Statex

    \Procedure{D$_{min}$-I}{$X$,$y$}
    \State $C_{Xy} \gets \infty $
    \ForAll {$u \in g.V$} \Comment{all nodes in $g$}
    \If {$y \in u $}
    \State $C_{Xy} \gets$ \Call{ShortestPath}{$X$,$u$} 
    \ForAll  {$v$ where $(u,v) \in g.E_2$}
    \State $C_{Xy} \gets C_{Xy} + $ \Call{D$_{min}$}{$v$, $final$}
    \EndFor
    \EndIf
    \EndFor
    \State $C_{Xy_{min}} \gets$ select the minimal $C_{Xy}$ for all $u$
    \State \Return $C_{Xy_{min}}$
    \EndProcedure

    \Statex

    \Procedure{D$_{min}$-II}{$X$,$y$}
    \ForAll {$u$ reachable from X via Type I edges and has outgoing Type II edge(s), $y \notin u $}
    \State $C_{Xu}\gets$ \Call{ShortestPath}{$X$,$u$}; $C_{vy}\gets \infty$
    \ForAll {$v$ where $(u,v) \in g.E_2$}
    \State $C_{vy} \gets $ \Call{D$_{min}$}{$v$,$y$}; $C_{tf}\gets 0$
    \ForAll {$t$ where $(u,t) \in g.E_2 \wedge t \neq v$}
    \State $C_{tf} \gets C_{tf}$ + \Call{D$_{min}$}{$t$, $final$}
    \EndFor
    \State  $C_{vy} \gets C_{vy} + C_{tf}$
    \EndFor
    \State $C_{vy_{min}} \gets$ select the minimal $C_{vy}$ for all $v$

    \State  $C_{Xy} \gets C_{Xu}$ + $C_{vy_{min}}$
    \EndFor
    \State $C_{Xy_{min}\gets}$ select the minimal $C_{Xy}$ for all $u$
    \State \Return $C_{Xy_{min}}$
    \EndProcedure
  \end{algorithmic}

\end{algorithm}

In Algorithm \ref{alg:shortest}, we show how to find a shortest derivation from $X$ to the target $y$. The output \Call{D$_{min}$}{$X$,$y$} records the {\it cost} of this derivation (the length of the string generated from the derivation). Once the derivation is found, we test if the generated string $\beta$ is the subsequence of a given $\alpha$ (see Definition 8). If yes, we report the solution; if not, we continue finding the next shortest derivations until $\beta$ is a subsequence of $\alpha$.

At line~4, we compare two cases: in \textit{D$_{min}$-I} at line~5, the target $y$ is reached through only Type~I edges and in  \textit{D$_{min}$-II} at line~14, $y$ is reached through Type I and Type~II edges. In the first case, at lines 7--8, we first find all the nodes that contain $y$. For each of such node $u$, we run a shortest path algorithm on the paths consisting of only Type I edges to find the minimal cost from $X$ to $u$ (line 9). For any Type II edges from $u$ (line 10), we compute the minimal cost from $v$ to the {\it final} node that only contains terminals (line 11). At line~12, we select the minimum for all such $u$.

In the second case, at line~16, we first find the minimal cost from $X$ to $u$, where $X$ and $u$ are reachable through only Type I edges, $u$ has outgoing Type II edge(s), and $y \notin u$. For any Type II edge $(u,v)$, we compute the minimum cost from $v$ to the target $y$ at line~18 and add the cost for the rest of the nodes connected from $u$ at lines~19--21. At lines~22 and 24, we select the best $v$ and $u$ respectively.

\noindent{\bf Example:} In Figure~\ref{fig:lazy-eval-example}, suppose we want to compute a shortest derivation from $X$ to $m$ containing Type II edges. Along $X\rightarrow uCDv$, $C_{Xu}$ at line~16 in Algorithm~\ref{alg:shortest} equals 2. At line~17, the two choices of $v$ are the nodes $C$ and $D$. Along $C\rightarrow m$, $C_{vy}$ at line~18 equals 1 and $C_{tf}$ at line~20 equals 2 (along $D\rightarrow nd$). On the other hand, when $v$ at line~17 is $D$, along $D\rightarrow gnh$ and $D\rightarrow nd$,  $C_{vy}$ at line~18 equals $\infty$, as $D$ cannot reach $m$. Therefore, $C_{vy_{min}}$ at line~22 equals 3, and  $C_{Xy}$ at line~23 equals 5. The loop at line~15 only iterates once, and we conclude $C_{Xy_{min}}=5$ at line~24.

\section{Building and Testing Code Fragments}~\label{sec:BuildandTest}
In this section, we present our approaches of resolving dependencies and testing code fragments.

\subsection{Resolving Dependencies for Code Fragments}~\label{sec:dep}
We developed two approaches to identify the dependencies needed to build code fragments. In the first approach, we process the project and store all its definitions into a database. When handling each segment from the project, we retrieve the definitions on demand and patch them based on their orders in the original project. This approach does not require to build the project; as long as its syntax errors do not affect the segment, we still can compile the segment.

In the second approach, \helium finds the header files that contain the symbol definitions needed by the segment based on where the segment is located in the original project. \helium then includes these header files in the code fragment and link the segment with the object files of the original project.

To generate an executable, we also developed two approaches to prepare a build script for the code fragment.  In the first approach, \helium uses the header files included in the code fragments to figure out which object files in the original project and which libraries should be linked to. In the second approach, we recorded all the compiler and linker flags used to build the original project via tools like  {\it bear}~\cite{bear-url} and the CMake's \textit{export} command. We then create a build script based on these flags.

\subsection{Testing with KLEE, Fuzzers and Valgrind}~\label{subsec:Testing}

\noindent{\bf Creating test harnesses:} \helium adds a {\tt main} function, which serves as the top level function for the code fragment. It also inserts the code that can supply generated test data to the input variables. We used the approach in~\cite{2009-ISSTA-Jiang-Automatic} and performed {\it def-use} analyses on the code fragments to identify the input variables: Variable $v$ is an {\it input variable} if there is no definition of $v$ found before a use of $v$ in the code fragment.

\noindent{\bf Generating test inputs:} If we can determine the value of a variable before the code fragment, we use this value to initialize the variable. Otherwise, we initialize input variables using automatically generated test inputs. We currently implemented random input generation that supports integers, floats, chars, arrays, pointers and the data structures composed with these types. We also applied Radamsa and KLEE to further generate test inputs. We selected Radamsa because it can use a random input as a seed to generate useful inputs, and it has successfully found many vulnerabilities in real-world software such as {\it chrome}, {\it firefox}, {\it php}, {\it tor}, {\it libxslt} and {\it vlc}. 

\noindent{\bf Constructing test oracles:} We identified two types of useful tests, {\it valid} and {\it pass} tests. A valid test case triggers the failure specified in our test oracles. A test oracle consists of the failure location and failure symptoms, extracted from the corresponding static analysis warning. During testing, we compare if the actual failure location is matched to the one specified in the oracle, similar to the approach in~\cite{2019-AutomatedCustomized}.  For failure symptoms, we developed a mapping from warning types to the failure types that Valgrind and KLEE can report. Some warning types cannot be directly linked to the symptoms reported by dynamic tools. We thus also add assertions to help with the match. For example, we add {\tt assert(false)} at the dead code; when the assertion is triggered, we confirm it as a false positive.  Valgrind and KLEE can both report assertion failures.

We analyzed the static warning types and classify them into two categories. Ideally, path-sensitive tools can report warning paths that contain all the statements of relevance to the bugs. In such cases, the {\it positive} warnings require only one valid test case to demonstrate the bug, e.g., buffer overflows and divide-by-zero. We confirm them as true positives. For such warnings, when testing never triggers the failures defined in the oracles, we report them as likely false positives. Developers can prioritize them accordingly. The {\it negative} warnings specify the bugs with ``the code cannot", e.g., unreachable conditions or dead code. For these warnings, we confirm them as false positives by showing one counter example through testing that such conditions can happen.

\section{The Scope and Limitations of Our Approach}~\label{sec:scope}
\helium provides a syntactic patch for a static warning path by analyzing the parse trees of the program. If the original warning path misses a data dependency, our syntactic patching algorithm may add it when fixing syntax errors, but it does not guarantee that all the data dependencies missed by static analyses are patched. When the necessary failure-inducing statements are missed in the warning paths, or when the paths are actually not reachable from the beginning of the program, i.e., dead code, \helium can report inaccurate results, manifesting as false positives and false negatives; however, from our experience with commercial static analysis tools, we have found that these tools work well with our assumptions. They are conservative and often provide more statements than necessary to help developers to diagnose the warnings.

One drawback of static analysis tools is that they can generate many false positive paths as a result of over-approximation. {\it Helium}, as a testing tool, is useful for distinguishing buggy paths from false positive paths. Importantly, using {\it Helium}, developers now can obtain the actual, failure-inducing input and executable, in addition to static paths, to diagnose the issue. Of course, there can be cases where \helium fails to find a failure-inducing input and thus fails to prioritize the true positive warnings. Also, there are types of bugs, such as concurrency bugs, that can be difficult for tests to manifest even when the warning is valid.

\section{Evaluation}
Our evaluation aims to answer the following questions:

\begin{itemize}
  \item {\bf RQ1.} How effectively can our LCA-based patching and dependency resolving techniques compile code fragments? 
  \item {\bf RQ2.} How effectively can we generate harnesses, inputs and oracles for testing code fragments? Can we preserve semantics of warnings to trigger their bugs in testing?
  \item {\bf RQ3.} How effectively can we validate real-world static warnings?
\end{itemize}

\subsection{Experimental Setup}~\label{subsec:ExpSetUp}

To answer the research questions, we implemented \textit{Helium} for C programs using Clang~\cite{2004-CGO-Lattner-LLVM,2008-BSD-Lattner-LLVM}, pycparser~\cite{repo-pycparser} and srcML~\cite{repo-srcml}. The LCA patching algorithm is implemented in {\it Racket}~\cite{repo-racket} and the testing framework is implemented in C++.
For building the code fragments, we tried the two approaches specified in Section~\ref{sec:dep}. We used the second approaches for both compiling and generating build scripts, as they were shown to better handle real-world code. We manually (one-time effort) created a mapping between static and dynamic warnings similar to~\cite{2019-AutomatedCustomized}. The mapping is implemented in a script to enable automatic matching for individual segments.

\noindent{\bf Static analysis tools:} We surveyed all the static analysis tools (for C/C++ code) listed on the NIST website~\footnote{https://samate.nist.gov/index.php/Source_Code_Security_Analyzers.html}. Among the 24 tools, all of the 8 commercial static analyzers~\cite{codesonar,polyspace,doulecheck,Klocwork,Ctest,Astree,coverity,SonarQube} report paths in the static warnings while for the 16 open-source tools, only {\it clang-analyzer} and {\it Oink-stack} provided paths. {\it Oink-stack} cannot be compiled, and {\it clang-analyzer} is not {\it CWE}~\footnote{https://nvd.nist.gov/vuln/categories} complete and reported very limited warnings for our benchmarks. To demonstrate that \helium can generally handle many categories of warnings, we used the two popular, CWE complete, commercial static analyzers, {\it PolySpace} and the {\it Commercial Tool} (anonymous for our license agreement) in our studies. 

\begin{table*}[ht]
  \centering
    \caption{Results of RQ1: Compiling code fragments~\label{tbl:rq1}}
  \begin{tabular}{l||c||c|c||c|c|c|c||c|c|c|c||c|c}
    \hline
      \multirow{2}{*}{Projects} &
      \multirow{2}{*}{Size(sloc)} &
      \multicolumn{2}{|c||}{Static Warnings} &
      \multicolumn{4}{|c||}{Syntactic Patching} & 
      \multicolumn{4}{|c||}{Dependencies} &
      \multicolumn{2}{|c}{Code Fragment Size}
      \tabularnewline  \cline{3-14}
                          &        & C    & P   & np & LCA & R & M & np & Helium & R & M & np    & LCA \tabularnewline
    \hline \hline
      polymorph-0.4.0     & 3.6k   & 6    & 7   & 0  & 13  & 0  & 0     & 0  & 12     & 0  & 0     & 21.9  & 41.4 \\\hline
      ncompress-4.2.4     & 6.7k   & 15   & 2   & 0  & 17  & 0  & 0     & 0  & 13     & 0  & 0     & 157.1 & 183.9 \\\hline
      man-1.5h1           & 26.9k  & 37   & 18  & 3  & 22  & 4  & 3     & 1  & 36     & 5  & 2     & 65.5  & 99.9 \\\hline
      gzip-1.2.4          & 35.3k  & 22   & 19  & 1  & 20  & 0  & 1     & 2  & 38     & 1  & 2     & 35.3  & 52.7 \\\hline
      bc-1.06             & 49.9k  & 6    & 45  & 1  & 4   & 0  & 1     & 1  & 41     & 2  & 1     & 72.8  & 75.3 \\\hline
      squid-2.3.STABLE5   & 329.9k & 118  & 85  & 2  & 95  & 1  & 7     & 5  & 187    & 1  & 4     & 43.9  & 61.9 \\\hline
      cvs-1.11.4          & 462.3k & 447  & 138 & 13 & 150 & 3  & 13    & 14 & 355    & 5  & 10    & 123.3 & 132.3 \\\hline
      httpd-2.0.48        & 724.4k & 94   & N/A & 1  & 1   & 1  & 1     & 7  & 61     & 6  & 0     & 61.1  & 84.6 \\\hline\hline
      findutils-e6680237  & 92.5k  & 92   & 116 & 6  & 28  & 7  & 9     & 11 & 106    & 6  & 11    & 65.2  & 89.2 \\\hline
      make-0afbbf85       & 130.4k & 91   & 26  & 6  & 0   & 3  & 6     & 5  & 77     & 1  & 3     & 78.1  & 86.7 \\\hline
      grep-c32c0421       & 407.0k & 79   & 91  & 1  & 91  & 3  & 3     & 3  & 124    & 3  & 4     & 53.9  & 76.7 \\\hline
      coreutils-0928c241  & 676.7k & 293  & 108 & 22 & 117 & 7  & 39    & 42 & 290    & 22 & 45    & 73.4  & 92.6 \\\hline\hline 
      summary & 2,942.6 k & 1300   & 655  & 56  & 618& 29  & 79 & 91    & 1340   & 52 & 82    & 70.9 & 89.8
      \\\hline
  \end{tabular}
  \end{table*}

\noindent{\bf Benchmarks:} We used C projects in CoreBench~\cite{bohme2014corebench} (we used the first listed buggy version of all 4 projects) and BugBench~\cite{2005-Bugbench} (it released 11 programs and 8 are C programs which we used). The 12 projects consist of 2.9 million lines of code~(sloc)~\footnote{the lines are calculated by {\it tokei}, https://github.com/XAMPPRocky/tokei}, shown in the first two columns of Table~\ref{tbl:rq1}. From \codesonar and {\it PolySpace}, we processed a total of 1955 warnings of 41 categories. We did not include code smell warnings that manifest no dynamic symptoms, and we also did not include the warnings that require dynamic-checks not yet supported by KLEE, Valgrind or our assertions.

\noindent{\bf Experimental design, metrics and baselines:} To answer RQ1, we used three metrics: 1) the number of code fragments successfully parsed under {\it Syntactic Patching} in Table~\ref{tbl:rq1} (for evaluating LCA algorithm), 2) the number of code fragments successfully compiled under {\it Dependencies} (for evaluating our dependency solver), and 3) the average size of code fragments patched under {\it Code Fragment Size} (for demonstrating LCA patch size). Here, we compared Helium with three baselines: Column {\it np} shows the results from the code fragments directly taken from static warnings without patching; Columns {\it R} and {\it M} report the results from RLAssist~\cite{2019-DeepReinforce} and MACER~\cite{2020-Macer}, the two state-of-the-art compiler error fixing tools, respectively. To avoid gathering results in favor of our tool, when computing compilation rate, we ran our dependency solver for each segment and then send it to RLAssist and MACER for patching. Hence, the data under {\it Syntactic Patching} and {\it Dependencies} are observed under the same input for all the four settings.

To answer RQ2, we report the number of executable test cases generated and also the number of test cases triggered by random testing, Radamsa and KLEE under Columns {\it Executable Tests}, {\it Random}, {\it Radamsa}, and {\it KLEE} in Table~\ref{tbl:dynTest}. We provide the total valid/pass test cases generated by the three approaches under Column {\it Summary of testing}. In addition to RLAssist and MACER (Columns {\it R} and {\it M}), we used {\it Unit Testing} as another baseline (Columns {\it U} in Table~\ref{tbl:dynTest}). Unit testing is a common industry practice, and combining static checking with unit testing has been shown useful for Java code~\cite{cnc}. It will be interesting to see how effectively unit testing can validate static warnings generated from the integrated C software. To construct a unit test, we took the functions where intraprocedural warnings are reported. We used Helium to resolve their dependencies and create test harnesses. 

In random testing, we ran 20 randomly generated test inputs for each {code fragment}. Initially, we ran 100 random inputs but found that the error is mostly triggered within the first 20 {inputs}. For Radamsa, we configured the fuzzing time as 15 minutes and used 100 random inputs as seeds. Similar to random inputs, we observed empirically that increasing the time further didn't improve the results. To ensure reproducibility, we ran the fuzzing twice and only reported the results that were consistent across the runs.

For RQ3, we use the approach in Section~\ref{subsec:Testing} to determine how many test cases confirm true positives and false positives (Columns {\it tp} and {\it fp} in Table~\ref{tbl:rq3}), and how many help increase our confidence that the warnings are likely false positives (Column {\it likely fp}). We assigned three authors to validate {a random selection of 50\% of} the results in Table~\ref{tbl:rq3}, following the literature~\cite{2019-DeepLearningBugs}. The three authors first independently validate the results for 5 code fragments and then met to discuss. After learning and agreeing on the inspection details, the authors independently inspected the other segments and compared the results.  If a disagreement occurs for a particular output, the authors discuss it until either the agreement is reached, or it would be listed as unknown.

We added two baselines to compare testing segments with testing entire programs. First, we ran {\it BovInspector}~\cite{2015-ASE-BovInspector}, the only tool we found that applied symbolic execution over entire software to trigger bugs, guided by the paths in static warnings. We also ran existing test suites shipped with the benchmarks using Valgrind and determine if any of the static warnings are triggered in testing.

\noindent{\bf Running the experiments:}  All of our experiments, except the training for RLAssist and MACER, were run on a VM with an 8 core Intel Haswell processor, 16GB memory and CentOS 8. We trained RLAssist and MACER using the hyperparameters recommended in~\cite{2019-DeepReinforce,2020-Macer}. Training RLAssist and Macer took 2.5 days and 15 minutes, respectively,  on an Intel Xeon Gold 6152 CPU @ 2.10GHz, with 128GB of RAM and 2x NVIDIA Tesla V100(32GB) GPUs.

\begin{table*}[ht]
\centering
\caption{Results of RQ2: Testing code fragments ~\label{tbl:dynTest}}
\resizebox{\textwidth}{!}{
    \begin{tabular}{l||c|c|c|c||c|c|c|c||c|c|c|c||c|c|c|c||c|c|c|c} \hline
      \multirow{2}{*}{Project} & 
      \multicolumn{4}{|c||}{Executable Tests}  &
      \multicolumn{4}{|c||}{Random (valid/pass)}  &
      \multicolumn{4}{|c||}{Radamsa (valid/pass)} &
      \multicolumn{4}{|c||}{KLEE (valid/pass)} &
      \multicolumn{4}{|c}{Summary of testing (valid/pass)}
      \tabularnewline  \cline{2-21}
                                                                & H    & U   & R  & M  & H      & U      & R    & M    & H      & U     & R   & M   & H      & U     & R    & M    & H        & U      & R    & M \\ \hline\hline
      polymorph-0.4.0                        & 9    & 0   & 0  & 0  & 0/4    & 0      & 0    & 0    & 0/4    & 0     & 0   & 0   & 0/5    & 0     & 0    & 0    & 0/13     & 0      & 0    & 0  \\\hline
      ncompress-4.2.4                        & 14   & 10  & 0  & 0  & 2/4    & 0/2    & 0    & 0    & 1/3    & 0     & 0   & 0   & 2/3    & 0     & 0    & 0    & 5/10     & 0/2    & 0    & 0  \\\hline
      man-1.5h1                              & 33   & 12  & 5  & 1  & 5/18   & 0/7    & 0/4  & 0/1  & 7/9    & 0     & 0/1 & 0   & 5/17   & 0/7   & 0/5  & 0/1  & 17/44    & 0/14   & 0/10 & 0/2  \\\hline
      gzip-1.2.4                             & 24   & 6   & 1  & 2  & 0/16   & 0/4    & 0/1  & 0/2  & 0/13   & 0/1   & 0   & 0   & 0/14   & 0/2   & 0    & 0/2  & 0/43     & 0/7    & 0/1  & 0/4  \\\hline
      bc-1.06                                & 37   & 1   & 0  & 0  & 0/23   & 0/1    & 0    & 0    & 1/22   & 0     & 0   & 0   & 0/22   & 0/1   & 0    & 0    & 1/67     & 0/2    & 0    & 0  \\\hline
      cvs-1.11.4                             & 312  & 61  & 5  & 7  & 9/158  & 3/37   & 0/4  & 0/6  & 10/51  & 3/3   & 0   & 0/1 & 7/126  & 3/15  & 0/3  & 0/5  & 26/335   & 9/55   & 0/7  & 0/12 \\\hline
      squid-2.3.STABLE5                      & 155  & 31  & 0  & 3  & 14/54  & 2/11   & 0    & 0/1  & 20/36  & 2/1   & 0   & 0   & 19/70  & 1/14  & 0    & 0    & 53/160   & 5/26   & 0    & 0/1  \\\hline
      httpd-2.0.48                           & 25   & 18  & 1  & 0  & 0/16   & 0/5    & 0/1  & 0    & 0/3    & 0/2   & 0   & 0   & 0/5    & 0/2   & 0/1  & 0    & 0/24     & 0/9    & 0/2  & 0  \\\hline\hline
      findutils-e6680237                     & 89   & 37  & 4  & 8  & 5/28   & 3/22   & 1/3  & 1/3  & 4/15   & 4/6   & 1/3 & 0/3 & 4/34   & 2/22  & 1/3  & 1/3  & 13/77    & 9/50   & 3/9  & 2/9  \\\hline
      make-0afbbf85                          & 64   & 14  & 0  & 2  & 3/31   & 0/10   & 0    & 0/2  & 2/11   & 0/2   & 0   & 0   & 1/24   & 0/5   & 0    & 0/2  & 6/66     & 0/17   & 0    & 0/4  \\\hline
      grep-c32c0421                          & 109  & 13  & 1  & 0  & 13/66  & 1/9    & 0    & 0    & 17/14  & 1/4   & 0   & 1/0 & 7/63   & 1/5   & 0/1  & 0    & 37/143   & 3/18   & 0/1  & 1/0  \\\hline
      coreutils-0928c241                     & 132  & 68  & 14 & 20 & 5/100  & 1/50   & 0/12 & 0/16 & 7/39   & 1/17  & 0/2 & 0/4 & 5/74   & 1/31  & 0/5  & 0/13 & 17/213   & 3/98   & 0/19 & 0/33 \\\hline\hline
      \multirow{2}{*}{Total}                               & 1003 & 271 & 31 & 43 & 56/518 & 10/158 & 1/25 & 1/31 & 69/220 & 11/36 & 1/6 & 1/8 & 50/457 & 8/104 & 1/18 & 1/26 & 175/1195 & 29/298 & 3/49 & 3/65 \\\cline{2-21}
     & \multicolumn{4}{|c||}{1348} & \multicolumn{4}{|c||}{68/732} & \multicolumn{4}{|c||}{82/270} & \multicolumn{4}{|c||}{60/605} & \multicolumn{4}{|c}{210/1607} \\\hline
    \end{tabular}
}
\end{table*}

\subsection{Results of RQ1}~\label{ex1}
In Table~\ref{tbl:rq1} in Columns {\it np} under {\it Syntactic Patching} and {\it Dependencies}, we show that only 56 out of 1955 (2.8\%) warnings can be directly parsed, and that only 91 (4.6\%) warnings are compiled with the dependencies Helium provided. The results confirmed our assumption that most of the warnings generated by static analysis tools cannot be compiled directly.

Applying LCA syntactic patching, we successfully parsed 618 (31.6\%) code fragments, shown under {\it Syntactic patching/LCA}, and after further resolving dependencies, we compiled a total of 1340 (68.5\%) code fragments, shown under {\it Dependency/Helium}. We tried three tools to collect parsing rates, {\it pycparser},  {\it Clang} and {\it GCC} with {\tt -fsyntax-only} flags. We found that all of the tools require some type and macro information to report successful parsing~\cite{manual-pycparser,manual-gcc,manual-clang}. As a result, after we provide dependency information, the compilation rate is significantly improved. We used {\it pycparser}, as it requires the least dependencies and more accurately reports the syntax errors than the other two parsing tools.

Our results show that \helium works consistently for all the warning types and projects we have experimented with. We inspected 20\% of the compilation failures and found that about 60\% of the failures are caused by incorrect compiler flags. We used one generic makefile per project to compile all the code fragments; however, for some projects, certain segments require additional compiler flags to handle macros and variable declarations.
About 40\% of failures are caused by bugs in srcML and also by the fact that our prototype does not yet support all of the C language’s features e.g. nested macros.

Helium outperformed both RLAssist and MACER significantly. Under Columns {\it R} and {\it M}, we show that the parsing rates for RLAssist and MACER are 29 (1.4\%) and 79 (4\%) respectively, and the compilation rates are 52 (2.6\%) and 82 (4.1\%) respectively. We observed that some warnings miss only ";" or "\}" where the two tools can help. However, the two tools are not effective for the majority of real-world static warnings, as we found that they mutated and deleted lines to fix compiler errors. We analyzed the patches produced by {\it Helium}, and found that similar to the types of constructs added in Figure~\ref{example}, the patches included the relevant control-dependent nodes and sometimes data-dependent nodes missed by static analysis tools.

Under {\it Code Fragment Size}, we show that the LCA patch size is small, 18.8 sloc on average, and the average patched code fragment is 89.8 sloc.  The patched programs generated by RLAssist and MACER mostly have the same sizes as unpatched versions.
We report an average of 0.7 s per segment for LCA patching and 16.8 s for generating a compilable unit from a warning. Our approach is scalable in that the time of processing a warning is independent of the size of the software project. Therefore, we can handle warnings from all of the real-world programs in our benchmarks, the biggest of which is 676.7 k sloc.

\subsection{Results for RQ2}~\label{ex2}
Helium successfully generated 1003 executable test cases, compared to 271, 31 and 43 for unit testing, RLassist and MACER respectively shown under {\it Executable Tests} in Table~\ref{tbl:dynTest}.
We were not able to generate executables for all the complied code fragments because of the link errors and the errors of initializing input variables in test harnesses, specifically,  (1) certain macros and function definitions can take different arguments and return types depending on the compiler flags, which we did not always set correctly; (2) some functions in the project are declared as {\it static} or {\it inline} and only can be linked by the functions in the same files but not by our code fragments; and (3) our prototype is not yet able to recognize constant variables and tried to initialize them with test inputs.

Among the tools, \helium triggered the most bugs and assertions and reported 175 valid test inputs, shown under {\textit{Summary of testing} Column {\it H}}. Unit testing, RLAssist and MACER reported {29, 3, and 3} valid tests respectively under {\it Summary of testing} Columns {\it U}, {\it R} and {\it M}.
\helium can handle interprocedural warnings (which are 59\% of the total warnings) that unit testing cannot handle. Even comparing only intraprocedural warnings, \helium triggered 33 of the 802 warnings while unit testing only triggered 11. This indicates that testing code fragments more effectively triggered the bugs and assertions than testing entire procedures. 

Specifically, among the total 802 intraprocedural warnings, \helium and the unit testing approach produced 235 common test cases, 6 of which are triggered by both approaches. This provides the evidences, besides our reasoning in Section~\ref{sec:semantic}, that \helium preserved the semantics of static warnings. Importantly, we have seen the advantages of testing code fragments over unit testing even for handling intraprocedural warnings. First, there are four cases, where unit testing failed to reach the bug due to the larger size of the function, but \helium triggered the bug. Second, there are 23 cases where \helium succeeded and unit testing failed, and 5 cases where unit testing succeeded but \helium failed, mostly due to the build errors. We can see that \helium doesn't yet support all the C features, which impacted both \helium and unit testing but unit testing is affected more and produced more unbuildable cases because of its additional code and complexity.

For RLAssist and MACER, we found that all the warnings triggered by the two tools are also trigged by {\it Helium}. Our inspection showed that RLAssist and MACER changed control flow of the segments and function signatures, which lead to low success rates. This further confirms the importance of preserving semantics during syntactic patching.

Comparing the three testing approaches in the row {\it Total} in Table~\ref{tbl:dynTest}, we found that Radamsa (fuzzing) performed the best for triggering bugs in static warnings, and reported 82 valid test cases, followed by random testing (generated 68 valid test cases), and then KLEE (60 valid test cases).  Radamsa reported the least number of pass test cases, as we observed that the fuzzer tried to crash our test harnesses. Radamsa and Random both triggered some unique warnings, but KLEE did not.

\subsection{Results for RQ3}~\label{ex3}
In Table~\ref{tbl:rq3}, we show that Helium significantly outperformed other baselines regarding the confirmed true positives (under Column {\it tp}), confirmed false positives (under {\it fp}), and likely false positives (under {\it likely fp}). We see that \helium confirmed 48 true positives, 27 false positives and 205 likely false positives. On the other hand, unit testing confirmed only 11 false positives and 35 likely false positives, and RLAssist and MACER both confirmed 1 false positive and 6 and 10 likely false positives respectively. Note that multiple test cases in Column {\it Summary of testing} in Table~\ref{tbl:dynTest} can be generated from a same warning by different tools. The numbers under {\it tp} and {\it fp} in Table~\ref{tbl:rq3} counted validated warnings, and thus are smaller.

{\it BovInspector} only finished running with {\tt make} and did not trigger any buffer overflows. The rest of benchmarks are either not terminated after 60 minutes or cannot be handled by KLEE. This indicates that testing code fragments can provide the scalability and practicality that cannot be achieved by performing symbolic execution on the entire software. 

We ran a total of 7748 existing tests found in our benchmarks, excluding {\tt polymorph}, {\tt gzip}, {\tt ncompress}, {\tt httpd}, {\tt man} and {\tt squid} in BugBench that do not have test suites. We matched 2 memory leak warnings for {\tt coreutils}.

\begin{table}[ht]
  \centering
  \caption{RQ3: Validating static warnings: Columns {\it tp}, {\it fp} and {\it likely fp} list the tools' output.~\label{tbl:rq3}}
  \begin{tabular}{l||c|c|c}
    \hline
    Tool          & tp & fp & likely fp\\\hline\hline
    Helium        & 48 & 27 & 205 \\\hline
    Unit testing     & 0  & 11 & 35 \\\hline
    RLAssist      & 0  & 1  & 6 \\\hline
    MACER        & 0  & 1  & 10 \\\hline
    BovInspector & 0  & 0  & 0 \\\hline
    Existing test suite   & 2  & 0  & 0 \\\hline
  \end{tabular}
\end{table}

One of the \helium true positives listed in Table~\ref{tbl:rq3} is matched to CVE-2001-1413, and the other one is matched to a real-world bug~\footnote{\url{https://bugzilla.redhat.com/show\_bug.cgi?id=40400}}. We also triggered two more real-world bugs documented in BugBench, after fixing some bugs in the test harnesses of two segments. All of the four matched real-world bugs are interprocedural bugs and are only triggered by \helium but not other tools.

We manually validated 50\% of true positives and false positives reported in Table~\ref{tbl:rq3} (process given in Section~\ref{subsec:ExpSetUp}). We found that 75\% of the reviewed results are indeed true and false positives as stated by 
{\it Helium}, among which, the correct confirmation rate for true positives is 83\%. Our further investigation shows that the imprecision is due to the incompleteness of the \helium implementation, e.g., we did not include a bounds-checking related to a nested macro (not yet supported by {\it Helium}) and triggered buffer overflow; our test harness failed to correctly initialize a structure variable used in an {\it Unreachable Call} warning.

\subsection{Threats to Validity}
To address the external threats to validity, we selected two popular commercial static analysis tools and processed close to 2000 warnings of 40+ types from 12 C projects and 2.9 million lines of code. To address the internal threat to validity, we inspected 20-30\% code fragments at each step of our implementation, and 50\% of final results to assure they are correct (following a protocol documented in~\cite{2019-DeepLearningBugs}). After validating our results, we believe that our success rates can be further improved after we handle more C features.  We used the models trained with the dataset shipped with RLAssist and MACER. The dataset may not represent the distributions of our code fragments.

\section{Related Work}~\label{related}
\noindent{\bf Automatically processing static warnings:} Our work is related to verifying and testing software based on static warnings~\cite{Muske2015EfficientEO,Nguyen2019ReducingFP,Zhang:2017:oopsla,parvez2016combining,li2013dynamically,li2017automatically,zhang2011combined}. Muske et al. added assertions for divide-by-zero and array out-of-bounds warnings and applied model checking to verify the assertions~\cite{Muske2015EfficientEO}. Parvez et al. and Zhang et al. used symbolic execution to trigger the warnings in binary applications~\cite{parvez2016combining,zhang2011combined}. Li et al. validated memory leaks using concolic testing along the paths reported in the warnings~\cite{li2017automatically,li2013dynamically}. Our work generated a small possible test case for each warning instead of testing entire software.

There are also approaches that identify patterns from warnings, source code and software repositories for predicting false positives~\cite{cheirdari2018analyzing,2019:koc:ICST,wang2018defect,ruthruff2008predicting,lee2019classifying,aman2019survival,cheirdari2018analyzing,yang2020recognize,alikhashashneh2018using}, and that use machine learning techniques to learn what are likely true and false positives~\cite{google,Flynn:2018,tripp2014aletheia,ruthruff2008predicting,lee2019classifying,alikhashashneh2018using,koc2017learning}. For example, Zhang et al. automatically learned and integrated the users' feedback to rank the warnings~\cite{Zhang:2017:oopsla}. Finally, there are tools that use multiple static tools to cross-validate and rank the warnings to increase their reliability~\cite{Flynn:2018,2017:xypolytos:QRSC}.

\noindent{\bf Auto-fixing compiler errors}: Existing techniques addressed sub-problems of compilation issues, such as resolving identifier names~\cite{2017-ASE-Zhong-Boosting,2016-ISSTA-Terragni-Csnippex}, inferring types~\cite{2008-OOPSLA-Dagenais-Enabling} and fixing parsing errors~\cite{1963-CACM-Error-Irons,1972-Journal-Aho-Minimum,1973-POPL-Graham-Practical,2019-GradingUncomp}. Recently, deep learning and reinforcement learning have been applied to fix compiler errors based on the supervised dataset and feedback of compiler warnings~\cite{2019-DeepReinforce,2017-DeepFix,2018-NeuroSym,2019-LearningLenient,2019-DeepDelta,2019-Graph2Diff,2019-CQI,2018-SytaxSens}. Automatic program repair tools also reported some successes to fix compiler errors~\cite{2019-Getafix,2019-iFixR}. These tools synthesized the patches based on bug reports or past fixes. Our approach analyzed parse trees for syntactic patching. It relies on neither supervised dataset nor sufficient tests and oracles for correctness.

\noindent{\bf Parse tree analysis:}
 There has been work on transforming parse trees to reduce the size of programs for debugging compilers. {\it Hierarchical delta debugging (HDD)} performed delta debugging on parse trees to generate smaller programs~\cite{2006-ICSE-Misherghi-Hdd}. {\it Generalized tree reduction}~\cite{2017-ASE-Herfert-Automatically} improves HDD and performs transformations on a sub-tree of the parse tree. These techniques used a search-based technique to transform parse trees, and are different from our syntactic patching algorithm.

\noindent{\bf Testing code fragments for other applications:} There have been research interests in analyzing and executing code fragments~\cite{2009-ISSTA-Jiang-Automatic,2013-Le-ICSE-Segmented,2016-ASE-Rodriguez-Cancio-Automatic,2008-FSE-Jiang-Profile, 2016-Onward-Zilberstein-Leveraging,2007-ICSE-Jiang-Deckard, Godefroid:2014:icse, deissenboeck2012challenges, satter2017similarity}. Godefroid proposed {\it micro execution}~\cite{Godefroid:2014:icse}, a VM technique for executing code fragments at the binary level.  EqMiner ran contiguous lines of code to detect semantic clones~\cite{2009-ISSTA-Jiang-Automatic}. Segmented symbolic analysis applied dynamic analysis on loops and library calls to help static analysis~\cite{2013-Le-ICSE-Segmented}. All the above work handled continuous lines of code. None have focused on systematically building and testing any code fragments. 

 \vspace{-0.1cm}
\section{Conclusions and Future Work}
This paper presents the techniques and a tool to validate static warnings. The key idea is to take the warning and patch it to generate semantically equivalent executable code fragments. We can then use existing testing tools to trigger the failures in the warnings. We formally defined what we mean by patching a code fragment and developed an algorithm to automatically generate such patches. We built a system that addressed the challenges of building and testing code segments. We achieved 68.5\% build rate for the complicated warnings reported by commercial tools. Our tool scales to all the real-world large C projects used in our evaluation. We confirmed the true positives that match to CVE and real-world bugs, in which other baseline methods did not succeed. In the future, we plan to further improve the robustness of our tool and explore more applications of testing code fragments. 

\section*{Acknowledgments}
We thank the anonymous reviewers for their valuable feedback. We thank Sebastian Elbaum for kindly shepherding our paper. We thank Ekene Shiafiwi Okeke for collecting static warnings from the {\it PolySpace} static analysis tool. This research is supported by the US National Science Foundation (NSF) under Award 1816352.

\bibliographystyle{ACM-Reference-Format}
\bibliography{helium}

\end{document}